%% file: tsg2022.tex
\documentclass[lettersize,journal]{IEEEtran}
\usepackage{amsmath,amsfonts,algorithmic,algorithm,array,textcomp,stfloats}
\usepackage{url,verbatim,graphicx,cite,color}
\usepackage[caption=false,font=footnotesize]{subfig}
\newcommand{\tcell}[1]{\begin{tabular}[x]{@{}l@{}}#1\end{tabular}}
\usepackage{lipsum,booktabs,multirow}

\makeatletter
\def\input@path{{./tables/}}
\makeatother
\graphicspath{{figures/}}

\usepackage{amssymb}

\newcommand{\Ua}[0]{\mathcal{U}_{\alpha}}
\newcommand{\UaFx}[0]{\mathcal{U}_{\alpha}^{\mathrm{Fix.}}}
\newcommand{\Ba}[0]{\mathcal{B}_{\alpha}}
\newcommand{\Ga}[0]{\mathcal{G}_{\alpha}}
\newcommand{\Ma}[0]{\mathcal{M}_{\alpha}}
\newcommand{\Oa}[0]{\Omega_{\alpha}}
\newcommand{\ccvpq}[0]{\mathrm{CCV}_{\mathrm{PQ}}}
\newcommand{\upf}[0]{\mathrm{CCV}_{\mathrm{UPF}}}
\newcommand{\statcom}[0]{\mathrm{CCV}_{\mathrm{STATCOM}}}
\newcommand{\mpt}[0]{\mathrm{CCV}_{\mathrm{MPT}}}

\newcommand{\DM}[1]{\textcolor{black}{#1}}

\begin{document}

\title{Multiplexing Power Converters for\\Cost-Effective and Flexible Soft Open Points}

\author{Matthew Deakin,~\IEEEmembership{Member,~IEEE}
        
\thanks{This work was supported by the Centre for Postdoctoral Development in Infrastructure, Cities and Energy (C-DICE) programme led by Loughborough University in partnership with Cranfield University and the University of Birmingham (C-DICE is funded by the Research England Development Fund); the EPSRC Multi-energy Control of Cyber-Physical Urban Energy Systems (MC2) project (EP/T021969/1), and the Royal Academy of Engineering under the Research Fellowship programme. Corresponding email: \texttt{matthew.deakin@newcastle.ac.uk}.}}


\maketitle

\begin{abstract}
The feasible set of real powers that can be transferred by a three-terminal Soft Open Point (SOP) can be increased by selecting non-uniform power ratings for each of the three ac/dc legs of the SOP, then connecting a multi-terminal switch (multiplexer) to the ac side of each of those converters to facilitate reconfiguration. This paper generalizes this concept, considering the real and reactive power that $n$ multiplexed ac/dc converters can transfer at an $m$-feeder bus. The performance of the device is studied numerically for a number of ac/dc sizing strategies through the volume of the feasible set of power transfers (the `capability chart volume', CCV) and distribution system loss reduction benefits (as an exemplar network service). Upper bounds on device performance are defined by considering the performance of a novel, idealised SOP consisting of a continuum of infinitesimal reconfigurable converters. Results demonstrate that the CCV can be more than doubled, with 99\% of the relative performance improvement of the idealised converter achieved with designs consisting of as few as four converters. SOP equipment costs reductions of 24\% are reported, with it concluded that reconfigurable, judiciously sized ac/dc legs can yield flexible and lower cost SOPs than conventional, hard-wired approaches.
\end{abstract}


\section{Introduction}
\IEEEPARstart{P}{ower} converters will be an integral part of future power systems, providing a flexible interface between power networks and low carbon technologies. Given their flexibility and controllability, technologies based on power converters can increase distribution network capacity, reduce system losses, or address issues with power quality. However, the unit cost of the semiconductor switches at the core of power converters remains high compared to classical power technologies \cite{huber2017applicability}, limiting their uptake. As a result, the past decade has seen growth in `semiconductor assisted' hybrid technologies that exploit the controllability of small, partially rated converters to dramatically improve the performance of ac systems whilst reducing the costs of necessary expensive semiconductor components. Systems making use of such an approach include Hybrid Power Transformers \cite{bala2012hybrid}, Hybrid On-load Tap Changers \cite{Rogers2014low}, or Doubly-Fed Induction Generators (DFIGs) \cite{abad2011doubly}. When these hybrid technologies combine excellent performance with cost-effectiveness, these hybrid systems are attractive for industry, leading to high market share \cite{yaramasu2015high}.

One power converter-based technology in which Distribution System Operators (DSOs) are showing rapidly growing interest is the Soft Open Point (SOP). This is an approach which advocates replacement of distribution system Normally Open Points (NOPs) with back-to-back ac/dc converters that allow flexible power transfer between adjacent feeders without needing meshed protection \cite{bloemink2010increasing,jiang2022overview}. Whilst most works have focused on developing sophisticated control strategies (e.g.,  \cite{sarantakos2022robust,yang2022cooperative}), the high costs of ac/dc converter capacity has motivated increasing research interest into approaches that increase the flexibility of SOPs or reduce the required ratings of the ac/dc power converters. `Phase changing SOPs' allow LV SOPs to connect the ac/dc converters from individual phases together in various permutations (e.g., Phase A transferring power to Phase B) to increase device flexibility \cite{lou2020new}. The Hybrid Open Point (HOP) proposes to install SOPs in parallel with ac switches, with so-called `Type 2' HOPs permitted to be installed normally closed points in the network to increase planning and operational flexibility. The use of Unified Power Flow Controllers for SOPs has also been explored in \cite{ukpn2018active}, reducing the necessary ac/dc converter capacity.

This work considers a different approach, proposing full reconfiguration capabilities for each SOP leg--i.e., enabling the ac side of each of the SOP's ac/dc converters to be dynamically linked to any of the feeders at the bus the SOP is installed at. There are a small number of works that consider a similar philosophy, such as the `MVAC switchyard' of \cite{majumder2021distribution}, although the crucial role of converter sizing that is studied here is not discussed. Similarly, in \cite{xi2021vehicle}, the authors describe how banks of modular vehicle-to-grid converters can be reconfigured through a `relay matrix' to flexibly transfer power from an electric vehicle to the grid or to the house. There are also works that consider low-level design of multi-terminal power converters (e.g., multiport converters, multi-source converters, and parallel interlinking converters), aiming to reduce device count or the required capacity of the semiconductor switches  \cite{tao2008multiport,najafzadeh2021recent}, although reconfiguration of converter ports, as part of the topology design, is not considered. In the author's previous work, it was shown that the feasible set of real powers that can be transferred between feeders in a three-terminal SOP can be increased not only by reconfiguration, but also by sizing those converters in a non-uniform way \cite{deakin2022design}.

To the best of the author's knowledge, there are no works that consider the benefits of reconfigurable ac/dc converters in the general case, particularly in the context of sizing strategies and how these influence the device capability chart and subsequent performance of the device (previously demonstrated to be a significant factor in \cite{deakin2022design}). This is a substantial gap as the network operator's primary interest for such a converter is in this capability chart, as it represents real and reactive power transfers that they can control to provide network services. Additionally, it is shown in this work that considering sizing enables a description of an \textit{idealised} design with continuous reconfiguration capabilities. It is also worth noting the timeliness of this gap as network operators explore new ways of delivering capacity to enable huge numbers of new connections of distributed low carbon technologies \cite{beis2022electricity}.

\begin{figure}\centering
\subfloat[An $m$-terminal conventional SOP]{\includegraphics[width=0.178\textwidth]{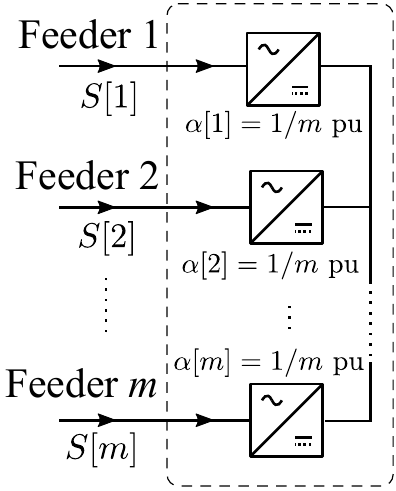}\label{f:sop_schematic}}
~~~~~
\subfloat[A multiplexed $ m $-terminal, $ n $-converter SOP (an $ (m,\,n) $ MOP)]{\includegraphics[width=0.245\textwidth]{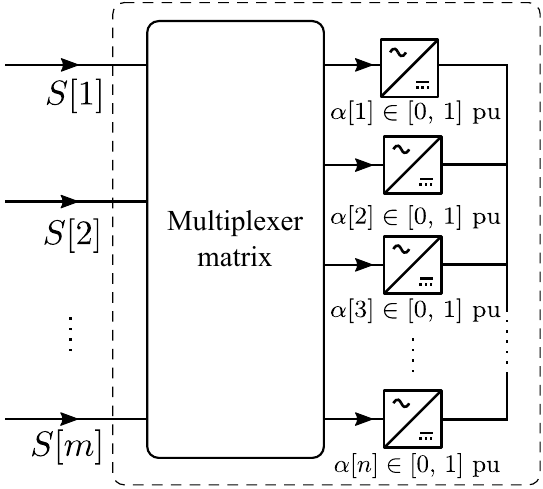}\label{f:mop_schematic_simple}}
\caption{A conventional $ m $-terminal SOP consists of $ m \geq 2 $ ac/dc converters, each with uniform capacity and each hard-wired to each of those $ m $ feeders at a node (a). In contrast, an $ (m,\,n) $ MOP consists of a number $ n \geq 2 $ ac/dc converters, sized in an arbitrary fashion, and each of which is connected on the ac side to a matrix of low-cost electromechanical multiplexers (b).}
\label{f:mop_vs_sop}
\end{figure}

The contribution of this work is to address this gap, describing and analyzing the generalized $ (m,\,n) $ Multiplexed Soft Open Point (MOP). This builds on the three terminal, three converter reconfigurable device described in \cite{deakin2022design}--the general MOP consists of an $m$ port device drawing real and reactive power through multi-terminal switches (multiplexers) to $ n $ non-uniformly sized ac/dc converters (Fig.~\ref{f:mop_vs_sop}). Analysis of the benefits of different ac/dc converter sizing strategies is considered through two intuitive and complementary methods, with converter sizing shown to be crucial for maximising device performance. The first method is to evaluate the hypervolume of the device's capability chart (the `Capability Chart Volume', CCV), considering all of the combinations of power transfers that are feasible for a given MOP design. Secondly, the performance of MOP designs embedded within distribution networks are considered for the loss reduction network service. For both metrics, the performance of a given MOP is compared against a hard-wired SOP and the idealised MOP. Numerical methods for evaluating both of these metrics are proposed, based on Monte Carlo Integration and Mixed-Integer Second Order Cone Programming (MI-SOCP) techniques respectively.

This paper is structured as follows. In Section~\ref{s:2_device_design}, Capability Chart Volumes are introduced as an intuitive method of evaluating MOP flexibility as a function of proposed sizing strategies, with the Idealised MOP defined formally to show how this can be used to bound converter performance. In Section~\ref{s:3_monte_carlo}, Monte Carlo Integration is proposed as a numerical method for systematically determining the CCV for designs with any number of power converters, which are required when analytic methods are not practical. Network-specific relative performance metrics are introduced in Section~\ref{s:4_network_performance}, based on characterizing a given MOP's performance in a distribution network as compared to both a conventional, hard-wired SOP and the Idealised MOP. CCVs are determined analytically and numerically in Section~\ref{s:5_cases} and compared against the network-aware performance criteria for a wide range of MOP designs, with a discussion considering further benefits and challenges of the MOP from the point of view of DSOs. Finally, salient conclusions are drawn in Section~\ref{s:6_conclusions}.

\section{MOP Design, Operation and Sizing}\label{s:2_device_design}

The MOP is designed to maximise the utilization of ac/dc converters within distribution systems by combining those converters with multi-terminal ac switches (multiplexers). In this section the Capability Chart Volume is introduced as a method for describing this increased flexibility. A set of four MOP `modes' are identified that highlight a range of possible end-uses of the MOP, each of which can be explored using the CCV method. Finally, sizing strategies for the MOP power converter are described, enabling both a description of an idealised MOP, and the systematic study of MOP performance as a function of the number of converters.

\subsection{MOP Operating Characteristics}

The premise of the MOP has been considered for increasing real power transfer in \cite{deakin2022design} for a 3-feeder, 3-terminal case, and can be summarised as follows. If 500 kW is to be transferred between any two feeders at a node with three feeders connected, then the conventional approach is to size three 500~kVA ac/dc converters of a SOP to enable that power transfer \cite{sun2020optimized,ji2017enhanced,lou2020new}. In total, this requires 1500~kVA of ac/dc power converter capacity. If instead one 500~kVA ac/dc converter and two 250~kVA ac/dc converters are connected, but each converter can connect to any of the three feeders at the node via a multiplexer (as in Fig.~\ref{f:mop_schematic_simple}), then 500~kVA can still be transferred between any two pairs of feeders. In addition, there is a locus of real power transfers that are feasible with the combinations of connections of those three converters amongst the three feeders. The multiplexed design therefore only needs 1000~kVA of ac/dc converters in total. Assuming that ac switches for reconfiguration have negligible cost compared to ac/dc converter capacity, then the cost of such a MOP will be reduced by a factor of 33\% as compared to the conventional SOP.

This paper considers the generalisation of this approach. Rather than considering only a 3-feeder, 3-converter case for real power transfer, the general case of an $m$-feeder, $n$-converter MOP is considered for both real power transfers and reactive power compensation.

\subsection{MOP Construction and Capability Charts}\label{ss:capability_charts}

\begin{figure}
\centering
\includegraphics[width=0.39\textwidth]{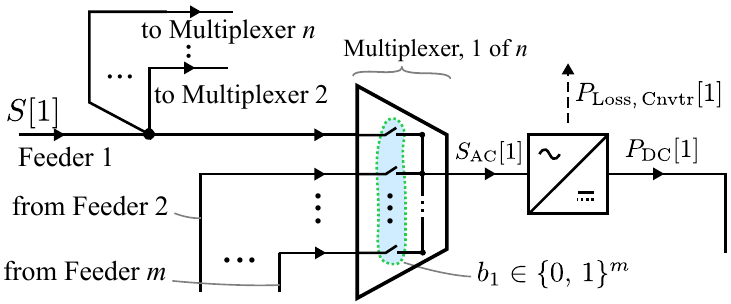}
\caption{The $j$th leg of an $m$-feeder, $n$-converter MOP, indicating the physical location of variables used for assessing both device-oriented and network-oriented performance. \DM{The multiplexer on each ac/dc converter (represented by a trapezoid) operates as a 1-to-$ m $ multi-terminal switch.} Modelling of the MOP capability can be achieved using a lossless model (Section~\ref{ss:capability_charts}) or with losses accounted for (Section~\ref{sss:mop_lossy_model})}
\label{f:mop_leg_tsg}
\end{figure}

A detailed model of the first leg of a MOP is shown in Fig.~\ref{f:mop_leg_tsg}, whilst the design of the MOP (i.e., per unit capacity of each of the $ n $ ac/dc converters) is defined by a vector $ \alpha \in \mathbb{R}^{n} $, as in Fig.~\ref{f:mop_schematic_simple}. \DM{The multiplexer element (Fig.~\ref{f:mop_leg_tsg}) is a 1-to-$ m $ switch, and is assumed to be low-cost as compared to the ac/dc legs. Each of the $ n $ multiplexers could be constructed of $ m $ parallel disconnect switches, a single $ m $-terminal electromechanical contactor, or from solid-state transfer switches.} \DM{The ac/dc converter (Fig.~\ref{f:mop_leg_tsg}) is assumed to be constructed using a voltage source converter topology, with this topology having advantages of flexible real and reactive power control, good voltage waveform control capabilities, and some fault isolation capabilities \cite{cao2016operating}.} 

Capability charts describe the set of all possible combinations of power transfers of a given MOP design, and can be defined as follows. Let the state of the $j$th multiplexer be described by $ b_{j}\in \{0,\,1\}^{m} $, with
\begin{equation}
\sum b_{j} = 1\,,\qquad j \in \mathbb{N}(n)\,,\label{e:relay_2}
\end{equation}
where $ \mathbb{N}(n) $ is the set of $n$ counting numbers, i.e., $ \mathbb{N}(n)=\{1,\,\cdots,\,n\}$. For example, a vector $b_{2}=[0,0,0,1]^{\intercal}$ implies that converter 2 is connected to the fourth feeder. 

Let $ B \in \{0,\,1\}^{m\times n}$ be the matrix that collects these vectors, i.e., $B=[b_{1},\,\cdots,\,b_{n}]$. The vector describing the converter capacity connected to each feeder, $ S^{+} $, is
\begin{equation}\label{e:conn_capacity_feeders}
S^{+} = B \alpha \,.
\end{equation}
The general capability chart (CC) and capability chart volume (CCV) can then be described as
\begin{align}\label{e:capability_chart}
\mathrm{CC} &= \left \{ S\,:\, \exists B \left [|S|\leq B\alpha \cap \sum P=0 \right ] \right \}\,,\\
\mathrm{CCV} &= \int _{\mathrm{CC}} d\mathbf{V} \,,\label{e:capability_chart_volume}
\end{align}
where $ S \in \mathbb{C}^{m} $ is a vector of complex power injections, $ P=\mathrm{Re}(S) $ is the vector of real power transfers, $ |\cdot| $ returns the elementwise absolute value of a given vector, and $ d\mathbf{V} $ is an incremental (hyper)volume. The two conditions of \eqref{e:capability_chart} ensure that the power injected from any feeder into the device is less than the capacity connected to a given feeder, and that the sum of power injections into the dc bus is zero (so that Kirchhoff's Current Law holds). \DM{Note that \eqref{e:capability_chart} assumes that losses in individual ac/dc converters are small as compared to the power transferred through them. This low loss assumption is typically valid, with losses through individual converters a few percent of transferred power \cite{cao2016benefits}. If very high accuracy is required for CCV calculations, a modified CCV can be determined by amending  \eqref{e:capability_chart} to consider a lossy converter model (as is necessary for system loss modelling, considered in Section~\ref{ss:formulation}).}

\subsubsection{CCV Modes}\label{ss:mop_use_cases} The CCV \eqref{e:capability_chart_volume} considers the magnitude, in a sense, of the device flexibility in real and reactive powers, and has units of kW$ ^{m-1} $kVAr$ ^{m} $. However, the MOP approach brings benefits even if a MOP is not required to utilise its full real and reactive capabilities. To explore this point, three further senses in which the device flexibility is affected by converter sizing strategies are considered in this work, to allow a more holistic exploration of device performance.
\begin{itemize}
\item Unity power factor power transfer mode ($\upf$, units: kW$ ^{m-1} $). This mode assumes that only real power transfers are required (i.e., $ Q=0 $), for example if networks have good power factor and no voltage congestion.

\item Static compensation mode, ($ \statcom $, units: kVAr$ ^{m} $). In this mode, it is assumed that the MOP is used to provide reactive power compensation only ($ P=0 $). This could be used to reduce losses, improve voltage regulation, or improve power quality.

\item Maximum power transfer mode ($ \mpt $, units: kW). The MOP is used for bulk real power transfer--only the maximum feeder-to-feeder power transfer is calculated.
\end{itemize}

Where it is potentially ambiguous, the CCV considering full real and reactive capabilities \eqref{e:capability_chart} is denoted $ \ccvpq $.

\subsection{MOP Converter Sizing}\label{ss:mop_sizing_strategies}

In general, the per-unit sizing of the MOP converters can have a substantial impact on feasible power transfers \cite{deakin2022design}. In contrast to that previous work, which explored only the sizing of three individual converters, this work considers \textit{strategies} for sizing the converters. These can be used to consider how changing the number of ac/dc converters $n$ and sizes $ \alpha \in \mathbb{R}^{n} $ change performance of the device. Three sizing strategies are considered in this work, as follows.
\begin{itemize}
\item The `Uniform' sizing strategy $ \Ua(n) $. Converters are all equally sized with rating $ 1/n $.
\item The `Bisection' sizing strategy $ \Ba(n) $. The first converter is sized at 0.5 pu; the next $n-2$ converters thereafter are sized with half of the size of the preceding converter. The $n$th converter is the size of the $n-1$th  converter.
\item The `Golden' sizing strategy, $ \Ga(n) $. The first converter is sized as 0.5 pu. The rest of the converters are then sized incrementally by splitting the smallest converter into two converters according to the golden ratio.
\end{itemize}
Finally, the conventional, hard-wired SOP (as shown in Fig.~\ref{f:sop_schematic}) is denoted as $ \UaFx(m) $. Note that per-unit sizing means that $ \sum \alpha = 1 $. 

\begin{figure}
\centering
\includegraphics[width=0.49\textwidth]{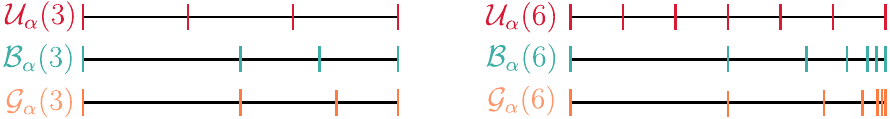}
\caption{A comparison of three MOP converter sizing strategies for converter numbers $n=3$ (left) and $n=6$ (right), with the width of the number line representing 1 pu ac/dc converter capacity, and coloured ticks the boundaries of the per-unit split of ac/dc converter sizes. $ \Ua(n),\,\Ba(n),\,\Ga(n)  $ represent the Uniform, Bisection and Golden sizing strategies respectively.}
\label{f:alpha_sizing_schemes}
\end{figure}

The three strategies are visualised on the unit numberline in Fig.~\ref{f:alpha_sizing_schemes}. It can be seen that the Bisection and Golden strategies consist of one large converter with progressively smaller converters, whilst the uniform sizing strategy naturally has converters that are uniform in size that get progressively smaller for larger $n$.

\subsubsection{The Idealised MOP}\label{sss:cardinal_mop}

The definition of the three sizing strategies in Section~\ref{ss:mop_sizing_strategies} admits the definition of the Idealised MOP design $ \Oa $. This design is ideal in the sense that it consists of an unbounded number of of infinitesimal ac/dc power converters, each with their own multiplexer connected to the $ m $ feeders, i.e.,
\begin{equation}\label{e:cardinal_mop_defn}
\Oa = \lim _{n \to \infty}\Ua(n) \,.
\end{equation}
Whilst such a sizing scheme cannot exist physically, the consideration of such a design has two interesting features that make it attractive for modelling. Firstly and most importantly, such a design gives an upper bound on the performance of any MOP. This is important as it allows converter performance to be upper bounded, so physically realisable designs can be compared, as is considered in Section~\ref{ss:opf_mop_benefits}.

Secondly, in introducing a continuum of power converters, the integer constraints required to define the CC in \eqref{e:capability_chart} are replaced with a continuous capacity constraint, resulting in feasibility problems which are simpler to solve. Similarly it is shown in Section~\ref{s:4_network_performance} that the Idealised MOP $ \Oa $ can be optimally scheduled using an SOCP rather than a mixed-integer SOCP for other MOP designs, resulting in a simpler optimization problem. The Idealised MOP is therefore attractive as a means of understanding how a MOP may fare as compared to a conventional fixed SOP without considering large numbers of MOP designs. 

\begin{figure}
\centering
\includegraphics[width=0.31\textwidth]{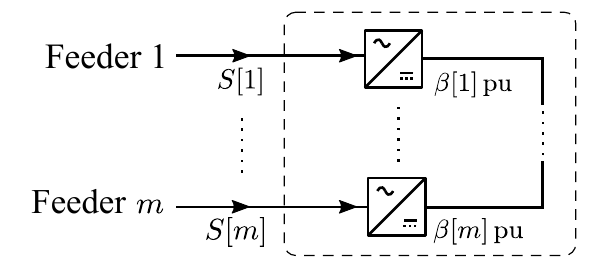}
\caption{\DM{The Idealised $ m $-terminal MOP $ \Oa $, defined as a limiting case of a Uniform sizing strategy \eqref{e:cardinal_mop_defn}, can be modelled as $ m $ ac/dc power converters, with $ \beta \in \mathbb{R}^{m}_{+}$ a vector collecting the (variable) capacity of each of these converters.}}
\label{f:mop_schematic_idealised}
\end{figure}

For example, the CC \eqref{e:capability_chart} for the Idealised MOP is instead
\begin{equation}\label{e:capability_chart_cardinal}
\mathrm{CC} = \left \{ S\,:\, \sum |S|\leq 1 \cap \sum P=0 \right \} \,,\\
\end{equation}
\DM{where the constraint $ |S|\leq B\alpha $ has been replaced with $ \sum |S| \leq 1 $. The latter is necessary and sufficient as any combination of capacities $ |S| $ whose sum is less than 1 pu can be created by allocating the $ i $th element of a vector of capacities $ \beta \in \mathbb{R}_{+}^{m} $ (as in Fig.~\ref{f:mop_schematic_idealised}) as the capacity connected to the $ i $th feeder, $ \beta[i]=|S[i]| $; conversely, any vector of apparent powers $ |S| $ whose sum is greater than 1 pu will be infeasible, as even the smallest capacity connected to each feeder (again as $ \beta [i]=|S[i]| $) will be insufficient to transfer the apparent powers $ S $.}

\subsubsection{Capability Chart Visualization} 
\begin{figure}\centering	
\subfloat[STATCOM capability charts]{\includegraphics[width=0.3\textwidth]{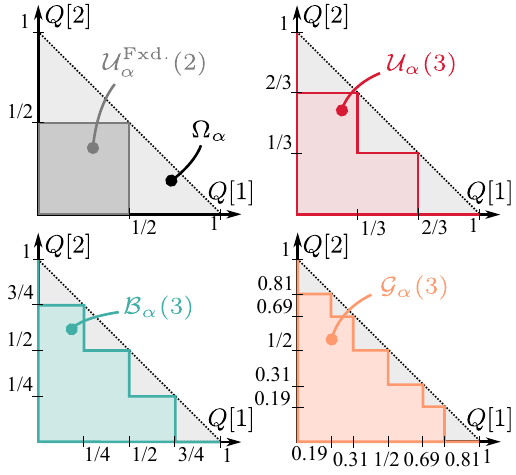}\label{f:cases_tsg}}
~
\subfloat[PQ capability charts]{\includegraphics[width=0.192\textwidth]{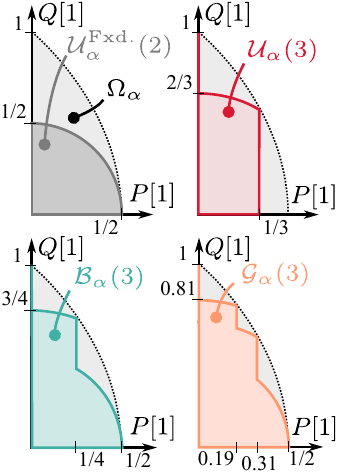}\label{f:cases_tsg_opf}}\\
\subfloat[\DM{UPF and MPT capability charts}]{\includegraphics[width=0.33\textwidth]{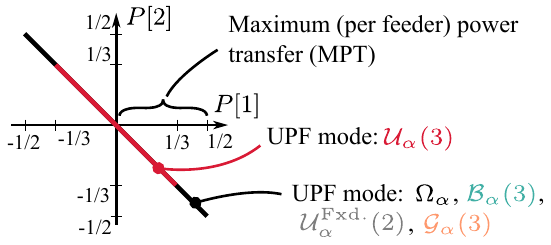}\label{f:cases_tsg_upf}}
\caption{Capability charts \eqref{e:capability_chart} (as described in Section~\ref{ss:mop_use_cases}) for five MOP designs and $ m=2 $ feeders. For the STATCOM mode, only the first quadrant is plotted; for PQ mode, the first quadrant is plotted, assuming the second feeder's $Q[2]=0$.}
\label{f:CCV_examples}
\end{figure}

It was noted in Section~\ref{ss:mop_use_cases} that the dimension of MOP capability charts increases with the number of feeders. In general, it is not possible to plot high-dimensional capability charts. However, when CCs are defined only in two dimensions they can be visualised \DM{(e.g., a range of two-dimensional, three-feeder UPF mode CCVs are plotted in \cite{deakin2023optimal})}. Fig.~\ref{f:CCV_examples} plots the \DM{$m=2$ feeder CCs for all four CCV modes. Note that, for the STATCOM and PQ CCs, only the first quadrant is plotted as the graphs are symmetric about the $ x $ and $ y $ axes.} It can be observed that the fixed, conventional SOP, denoted $ \UaFx(2) $, has a much smaller CC (and therefore CCV) than the other MOP designs \DM{for the STATCOM and PQ modes (the derivation of the PQ mode CCV in the Appendix is for all $ P$ and $ Q $; this two dimensional plot only shows $ Q[2]=0 $). In contrast, for MPT and UPF modes, for the two-feeder case, only the 3-converter uniform converter $ \Ua{(3)} $ has a reduced CCV as compared to the idealised MOP $ \Oa $.} The CCVs for each of the designs plotted in Fig.~\ref{f:CCV_examples} are determined analytically in Section~\ref{sss:results_analytic_ccvs}.

\section{Monte Carlo Integration for Evaluating Capability Chart Volumes }\label{s:3_monte_carlo}

In the previous section, CCVs were introduced as a means of quantifying the flexibility of MOP designs, with four modes of operation considered. In this section, the use of Monte Carlo Integration for calculating the CCV is explored, allowing network-agnostic comparison of MOP designs.

The evaluation of CCVs requires the assessment of multidimensional integrals \eqref{e:capability_chart_volume}. In low-dimensional cases, analytic results can be determined (see, e.g., Section~\ref{ss:case_ccvs}), but in the general case numerical methods must be used \cite{press2007numerical}[Ch.~4]. For this work, Monte Carlo Integration is used to assess these more complex integrals \cite{press2007numerical}[Ch.~7.7], having advantages of simple implementation and well-known convergence properties. 

The Monte Carlo Integration method can be summarised as follows. A sample space is defined with volume $ V $ that contains within it all feasible points within the integrand. $ N_{\mathrm{Total}} $ points are uniformly sampled within this space, and at each point a routine is run to evaluate if a given point is within the set \eqref{e:capability_chart}. The estimate of the integrand is then simply the ratio of the number of feasible points $ N_{\mathrm{Feasible}} $ to the total number $ N_{\mathrm{Total}} $ multiplied by sample space volume $ V $.

In other words, the estimate of the CCV, $ \overline{\mathrm{CCV}} $, is given by 
\begin{equation}\label{e:ccv_est}
\overline{\mathrm{CCV}} = V f\,, \quad f = \dfrac{N_{\mathrm{Feasible}} }{ N_{\mathrm{Total}} } \,.
\end{equation}
This method also gives an estimate of the standard deviation of the CCV estimate, $ \overline{\sigma}_{\mathrm{CCV}} $, as
\begin{equation}\label{e:ccv_sigma_est}
\overline{\sigma}_{\mathrm{CCV}} = V\dfrac{\sqrt{f(1-f)}}{\sqrt{N_{\mathrm{Total}}}}\,,
\end{equation}
which are used to estimate confidence intervals for the CCV. 

For this work, a (hyper)cube sample space $V$ is used. The feasibility problem that defines the CC \eqref{e:capability_chart} is a mixed-integer linear feasibility problem, which can be solved using standard optimization solvers.

\section{System-Oriented Performance Evaluation using Relative OPF Cost Reduction}\label{s:4_network_performance}

CCVs are an attractive method for evaluating MOP capabilities as they are simple and intuitive to describe, and they do not require assumptions about MOP behaviour within a network. If, however, a DSO does have a candidate network in mind for a potential investment in a MOP, then the known characteristics of that network can be used to provide more specific and accurate metrics describing the benefits and performance of a MOP design.

This approach is considered in this section, providing a description of the relative network benefits and performance. The latter makes use of the Idealised MOP as an upper bound on MOP performance, with the conventional hard-wired SOP used as the counterfactual. A MI-SOCP is proposed to enable the calculation of these metrics.

\subsection{Relative OPF Cost Reduction}\label{ss:opf_mop_benefits}

The installation of a MOP within an electrical distribution network allows the DSO (or other asset operator) to control real and reactive powers injected into the device. This means that the operational costs of the network $ f_{\mathrm{OPF}}$ can be reduced (e.g., for reduction of losses or minimization of solar curtailment). Assuming the MOP is the only asset that a DSO can control (so there is no time coupling), the minimum value of this cost function, $ f_{\mathrm{OPF}}^{*} $ at time $ t $ can be represented by
\begin{subequations}
\begin{align}\label{e:opt_value}
f_{\mathrm{OPF}}^{*}(\Ma,\,t) &= \min_{S} f_{\mathrm{OPF}} \\
\mathrm{Subject\,to} \quad S &\in \mathrm{CC}(\Ma) \\
F(S,\,t) &\in C\,, \label{e:congestion_constraints}
\end{align}
\end{subequations}
where $ \mathrm{CC}(\Ma) $ is the capability chart \eqref{e:capability_chart} of design $ \Ma $, and \eqref{e:congestion_constraints} represents congestion, with $ F $ the function mapping MOP injections $ S $ to current and thermal constraints. The case studies of this work have no congestion, so \eqref{e:congestion_constraints} trivially holds and so is not considered in the rest of this work.

Let the total OPF benefit for a design $ \Ma $ over $ T $ time periods be denoted by $ g^{*} $,
\begin{equation}\label{e:tot_opf_benefit}
g^{*}(\Ma) = \sum_{t=1}^{T}  f_{\mathrm{OPF}}^{*}(\Ma,t)\,.
\end{equation}
The relative network benefit $ \mu $ and performance $ \eta $ are
\begin{align}
\mu(\Ma) &= \dfrac{g^{*}(\Ma) - g^{*}(\UaFx) }{ g^{*}(\UaFx) } \,, \label{e:rel_ben_defn} \\
\eta(\Ma)  &= \dfrac{ g^{*}(\Ma) - g^{*}(\UaFx) }{ g^{*}(\Oa) - g^{*}(\UaFx) }\,. \label{e:rel_perf_defn}
\end{align}
The relative benefit $ \mu $ measures the relative increase in benefits due to the chosen design $ \Ma $ as compared to the conventional SOP $ \UaFx $. The relative performance $ \eta $ indicates how closely the design's performance is with respect to the performance of the Idealised design $ \Oa $ (e.g., a value of $ \eta = 80\%$ indicates a proposed design $ \Ma $ captures 80\% of the benefits of the Idealised MOP). Taken together, these metrics describe if a MOP provides a substantial benefit for a given network as compared to a conventional SOP, and what fraction of the potential benefit they can provide is.

\subsection{OPF for System Loss Reduction}\label{ss:formulation}

In this work, it is assumed that the DSO operates the MOP to reduce system losses $ P_{\mathrm{Loss}}^{\mathrm{Tot}} $, consisting of the sum of network losses, $P_{\mathrm{Loss}}^{\mathrm{Ntwk}} \in \mathbb{R}$, and converter losses $ P_{\mathrm{Loss}}^{\mathrm{Cnvtr}} \in \mathbb{R}^{n} $,
\begin{equation}\label{e:obj_val}
P_{\mathrm{Loss}}^{\mathrm{Tot}} = P_{\mathrm{Loss}}^{\mathrm{Ntwk}} + \sum_{i=1}^{n }P_{\mathrm{Loss}}^{\mathrm{Cnvtr}}[i]\,.
\end{equation}
Necessary converter and network models are described in the sequel to enable the reduction of these total system losses. \DM{In contrast to CCVs (Section~\ref{ss:capability_charts}), note that converter losses are modelled explicitly within the cost function \eqref{e:obj_val}. This is because the changes in converter losses are significant as compared to the changes in network losses, and so cannot be neglected \cite{cao2016benefits,sarantakos2022robust}.} Also note that, as in Section~\ref{ss:capability_charts}, all powers are in per-unit on the total ac/dc converter capacity base.

\subsubsection{MOP Model with Losses}\label{sss:mop_lossy_model}

Fig.~\ref{f:mop_leg_tsg} is a single line diagram of one leg of a MOP. \DM{As considered in previous works \cite{jiang2022overview,deakin2022design,deakin2023comparative}, the model used for the ac/dc converter (voltage source converter) has a constant apparent power limit (as in \eqref{e:capability_chart}) with losses modelled as a linear function of apparent power. That is,} the power losses $ P_{\mathrm{Loss}}^{\mathrm{Cnvtr}}[j] \in \mathbb{R}^{+} $ of the $j$th converter being proportional to the ac-side apparent power, i.e.,
\begin{equation}\label{e:conv_loss}
P_{\mathrm{Loss}}^{\mathrm{Cnvtr}}[j] = \kappa | S_{\mathrm{AC}}[j]| \quad \forall j \in \mathbb{N}(n) \,,
\end{equation}
where $\kappa$ is the converter loss coefficient. The real power balance across the $i$th ac/dc converter is
\begin{equation}\label{e:nodal_acdc}
P_{\mathrm{DC}}[j] + P_{\mathrm{Loss}}^{\mathrm{Cnvtr}}[j] = P_{\mathrm{AC}}[j] \quad \forall j \in \mathbb{N}(n) \,.
\end{equation}
As the converter application considered here is loss reduction, a relatively high ac/dc/ac efficiency of 98\% is considered following \cite{deakin2022design}, although it is worth noting that design of high efficiency converters is challenging \cite{huber2017applicability}.

The lossless CC defined in \eqref{e:capability_chart} must be amended to account for these converter losses. For the purposes of the optimal MOP scheduling problem, only the aggregate capacity of the converters connected to each feeder $ S^{+} $ is needed (as in \eqref{e:conn_capacity_feeders}), as the converter injections $ S_{\mathrm{AC}} $ can be determined after the optimal feeder injections $ S $ has been found (by splitting each power in proportion to the capacity of each converter). For example if 0.3 pu is drawn through feeder 1 connected to two ac/dc converters of capacity 0.15 pu and 0.3 pu, these two converters would draw 0.1 pu and 0.2 pu respectively. In contrast to the optimization proposed in \cite{deakin2022design}, this aggregation approach reduces the number of auxiliary variables in the optimization problem substantially.

Therefore, \eqref{e:conv_loss}, \eqref{e:nodal_acdc} are replaced with
\begin{align}\label{e:conv_loss_e}
\hat{P}_{\mathrm{Loss}}^{\mathrm{Cnvtr}}[i] &= \kappa |S[i]| \quad \forall i \in \mathbb{N}(m) \,,\\
\hat{P}_{\mathrm{DC}}[i] + \hat{P}_{\mathrm{Loss}}^{\mathrm{Cnvtr}}[i] &= P[i] \quad \forall i \in \mathbb{N}(m) \,.
\end{align}
where $ \hat{(\cdot)} $ indicates an the equivalent value taken by considering all converters connected to a feeder as one (as compared to the per-converter values shown in Fig.~\ref{f:mop_leg_tsg}). As in the CC \eqref{e:capability_chart}, the real power at the dc node must balance and the apparent power must be limited,
\begin{align}\label{e:nodal_dc_mod}
\sum _{i=1}^{m} \hat{P}_{\mathrm{DC}}[i] = 0&\,, \\ 
|S|\leq B\alpha\,,\qquad \sum B[:,j] =1 & \quad \forall j\in \mathbb{N}(n) \,. \label{e:repeat_s_ba}
\end{align}
The link between real, reactive and apparent powers (for \eqref{e:conv_loss}, \eqref{e:repeat_s_ba}) is made through the SOC relaxation
\begin{equation}\label{e:apparent_relaxation}
|S[i]| \geq \sqrt{P[i]^{2} + Q[i]^{2}} \quad \forall i \in \mathbb{N}(m) \,.
\end{equation}
Note that in this formulation the only place in which the problem increases with number of converters $ n $ is in the matrix $B$ linking converter apparent powers to power injections \eqref{e:repeat_s_ba}.

Finally, when modelling with the Idealised MOP $ \Oa $, the same loss model \eqref{e:conv_loss} is used. However, in this case, the constraint linking apparent power injections $S$ and converter capacities through $ B\alpha $ can be replaced with the continuous constraint using a continuous vector $ \beta \in \mathbb{R}^{m} $,
\begin{equation}\label{e:cardinal_constraint}
|S| \leq \beta \,, \quad \sum \beta=1\,, \quad |\beta|\geq 0 \,.
\end{equation}

\subsubsection{System Modelling}
Distribution network losses are modelled as a quadratic function in injections $ S_{\mathrm{inj}} \in \mathbb{R}^{2m} $ as
\begin{equation}\label{e:ntwk_loss}
P_{\mathrm{Loss}}^{\mathrm{Ntwk}} = S_{\mathrm{inj}}^{\intercal}Q_{\mathrm{Ntwk}}S_{\mathrm{inj}} + q_{\mathrm{Ntwk}}^{\intercal}S_{\mathrm{inj}} + c_{\mathrm{Ntwk}},\,
\end{equation}
where $Q_{\mathrm{Ntwk}},\,q_{\mathrm{Ntwk}},\,c_{\mathrm{Ntwk}}$ are the quadratic model coefficients, the first $m$ elements of $S_{\mathrm{inj}}$ the real part of $S[i]$, and elements $m+1$ to $2m$ of $S_{\mathrm{inj}}$ the imaginary part of $S[i]$ (i.e., $S_{\mathrm{inj}} = [\mathrm{Re}(S^{\intercal}),\,\mathrm{Im}(S^{\intercal})]^{\intercal}$). This approach is described in \cite{deakin2022design} and so is briefly summarised here. Firstly, the sensitivity matrix (Jacobian) of complex voltages in complex power injections is determined at the nominal power flow solution (see, e.g., \cite{bernstein2017linear}). Primitive admittance matrices are then used to determine a quadratic loss for each element, which are then summed to create the overall model \eqref{e:ntwk_loss}.

Together, the MOP optimal schedule is determined through the optimization problem
\begin{subequations}
\begin{align}
\min \: & \eqref{e:obj_val}\,, \label{e:opt_of} \\
\mathrm{s.t.} \: \eqref{e:conv_loss_e}  &- \eqref{e:apparent_relaxation} \,. \label{e:opt_cons}
\end{align}
\end{subequations}
To optimize for the Idealised MOP, the same formulation is used, with the exception that \eqref{e:repeat_s_ba} is replaced with \eqref{e:cardinal_constraint}.

\section{Case Studies}\label{s:5_cases}

In the previous two sections, we described approaches for evaluating the network-agnostic CCV (Section~\ref{s:3_monte_carlo}) and the network-aware relative performance and relative benefits (Section~\ref{s:4_network_performance}). In this section, we consider both of these approaches to explore how the MOP compares to conventional SOP designs, as highlighted via the proposed metrics.

Firstly, CCVs are calculated for each of the proposed operational modes (Section~\ref{ss:case_ccvs}). Analytical results are presented for a set of tractable cases, with numerical results reported for cases with larger numbers of converters and feeders. Relative benefits and performance are then calculated for three case studies (Section~\ref{ss:case_simulations}). Further challenges and opportunities for the MOP as compared to the conventional SOP for real-world implementation in distribution systems are then discussed in Section~\ref{ss:case_discussion}.

\DM{The optimization and feasibility problems can be solved by any modern solver with mixed-integer conic optimization capabilities}--the Mosek Fusion API \cite{mosek2021mosek} was used in this work to solve all of these problems. The conic relaxations in apparent power \eqref{e:apparent_relaxation} and for network losses \eqref{e:ntwk_loss} were found to be numerically exact to within a relative error of $ 10^{-4}$ across all simulations. Likewise, the maximum absolute mixed-integer optimization gap was 2.1~W, and the maximum relative difference between the losses obtained from the true non-linear solution (from OpenDSS) and the MI-SOCP was 2.45\%. Together, these indices indicate very good numerical performance of the proposed computational approaches.

\subsection{Evaluating Capability Chart Volumes}\label{ss:case_ccvs}

We first explore the CCVs using the $m=2$ feeder, $n=3$ converter case, as this problem admits analytic results for each of the CCV modes described in Section~\ref{ss:mop_use_cases}. We then explore CCVs numerically using Monte Carlo Integration for the unity power factor CCV, to consider the impact of increasing numbers of converters $n$ on CCVs.

\subsubsection{Two-Feeder Analytic CCVs}\label{sss:results_analytic_ccvs}
Table~\ref{t:analytic2feeder} reports the CCVs, determined analytically, for five (2, 3) MOP designs: the fixed conventional SOP $ \UaFx $, the $n=3$ converter designs of the Uniform $ \Ua $, Bisection $ \Ba $ and Golden $ \Ga $ sizing strategies, and finally the Idealised design $ \Oa $. For reference, Fig.~\ref{f:alpha_sizing_schemes} plots the relative sizes $ \alpha $ for each the three sizing strategies.

\begin{table}
\input{analytic2feeder}
\label{t:analytic2feeder}
\end{table}

The value of $ \mpt $ is given by the size of the largest converter. The value of $ \upf $ is determined by considering these maximum points on the line $ P[1]=P[2] $ (e.g., from ($-0.5$, $-0.5$) to (0.5, 0.5) for the conventional SOP $ \UaFx $). The value of $ \statcom $ can be calculated by plotting and calculating the area of the capability charts in the plane--the first quadrant of all of the CCs in reactive power are plotted in Fig.~\ref{f:cases_tsg} \DM{(note that the capability charts in Fig.~\ref{f:cases_tsg} are symmetric about the $ x $ and $ y $ axes).} Finally, it is shown in Appendix that the PQ capability $ \ccvpq $ can be evaluated for four of the five cases using elliptic integrals, whilst the value of $ \ccvpq $ for the Idealised MOP $ \Oa $ is evaluated using Simpson's rule.

A number of observations can be drawn from Table~\ref{t:analytic2feeder}. Firstly, it can be seen that CCVs do not necessarily increase with the number of converters $n$--the uniform design $ \Ua(3) $ can only transfer 1/3~pu, and so the $ \mpt $ is 1/3. However, this design does still outperform the conventional SOP $ \UaFx $ for both $ \statcom $ and $ \ccvpq $, and so could prove to be more effective than a SOP when both real and reactive power is required.

In contrast to the Uniform sizing strategy $ \Ua $, the Bisection $ \Ba $ and Golden $ \Ga $ strategies both maintain the maximum possible values of $ \mpt$ and $\upf $, but also increase $ \statcom $ and $ \ccvpq $. The Golden strategy $ \Ga $ has higher values of both STATCOM and PQ CCVs than the Bisection method $ \Ba $, and so it might be expected that such a design would perform better in practise.

Finally, as with the Bisection and Golden designs, the Idealised MOP $ \Oa $ cannot increase the values of $ \mpt $ or $ \upf $ compared to the conventional SOP $ \UaFx $, but this design does double the the value of $ \statcom $ and increase the $ \ccvpq $ by 85\%. It can be seen that the $ \Ga(3) $ design captures more than half of the potential increase in $ \statcom $ and $ \ccvpq $.

\subsubsection{CCVs with many converters}

\begin{figure}
\centering
\includegraphics[width=0.45\textwidth]{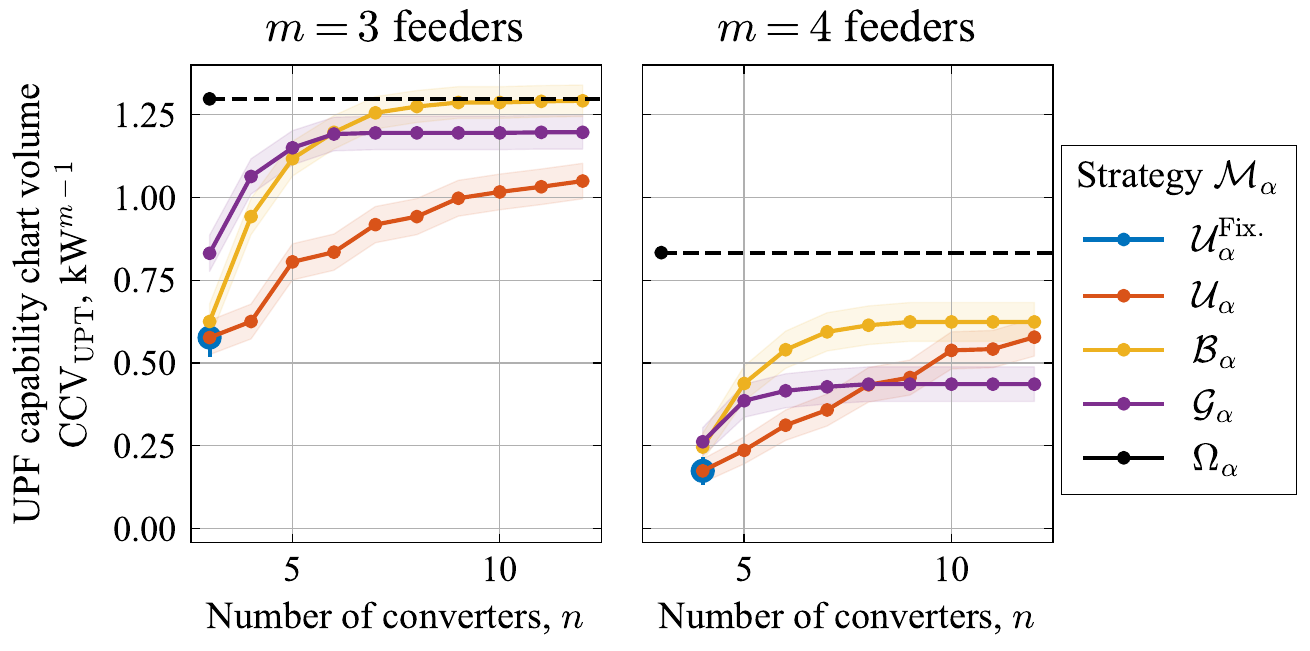}
\caption{Comparing the unity power factor volume $ \upf $ against strategy $ \Ma $ and number of converters $ n $. The solid line indicating the estimate of the CCV (from \eqref{e:ccv_est}), and the shaded area the 95\% region. The Idealised MOP $ \Oa $ can have a CCV many times that of the fixed MOP $ \Ua $.}
\label{f:pltUpfCaseStudies}
\end{figure}

When there are many converters, the complexity of the integrals necessitates the use of the Monte Carlo Integration, as described in Section~\ref{s:3_monte_carlo}. Fig.~\ref{f:pltUpfCaseStudies} plots unity power factor volume $ \upf $ for the three MOP sizing strategies, as well as the conventional SOP $ \UaFx $ and Idealised MOP $ \Oa $. On this figure, the shaded area represents the 95\% confidence interval derived from the standard deviation $ \overline{\sigma}_{\mathrm{CCV}} $ (from \eqref{e:ccv_sigma_est}), based on $ N_{\mathrm{Total}}=1000 $ function evaluations.

From this figure it can be observed that, as compared to systems with $m=2$ feeders, there is a much larger opportunity to increase the value of $ \upf $--the Idealised MOP $ \Oa $ more than doubles $ \upf $ compared to the conventional SOP $ \UaFx $ in the $m=3$ feeder case, and more than quadruples $ \upf $ in the $m=4$ feeder case.

For both $m=3$ and $m=4$ feeders, the Golden sizing $ \Ga $ strategy performs best for small numbers of converters $n$, as in the $m=2$ feeder case. However, for both 3 or 4 feeders, the Bisection sizing $ \Ba $ has a higher value of $ \upf $ as the number of converters $n$ increases. In both cases, $ \upf(\Ga(n)) $ appears to reach a plateau when there are $ n=8 $ converters. For the Bisection method $ \Ba $ and $ m=3 $ feeders, the value of $ \upf $ appears to approach the value realised by the Idealised MOP $ \Oa $ as $n$ increase above 10. For the case of $ m=4 $ feeders, it appears to plateau at a level close to three quarters of $ \upf(\Oa) $.

For both $ m=3 $ and $ m=4 $ feeders the Uniform sizing starts with a value of $ \upf $ that is substantially smaller than either the Golden or Bisection sizing strategies. However, by definition, the Uniform strategy is guaranteed to reach CCV of the Idealised MOP as $n\to\infty$ (by \eqref{e:cardinal_mop_defn}).

\subsection{Network Benefits and MOP Performance}\label{ss:case_simulations}

\begin{figure}\centering
\subfloat[33 Bus]{\includegraphics[width=0.132\textwidth]{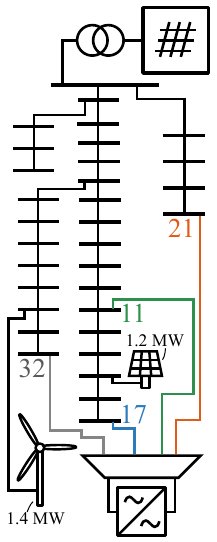}\label{f:case_33bus_4mop}}
~
\subfloat[GDS HVUG]{\includegraphics[width=0.135\textwidth]{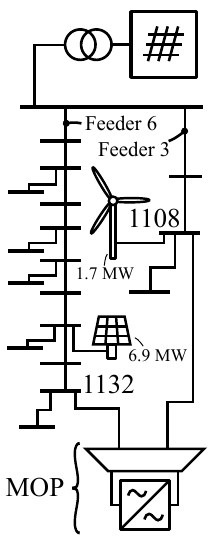}\label{f:case_study_hvug}}
~
\subfloat[Profiles]{\raisebox{1em}{\includegraphics[width=0.18\textwidth]{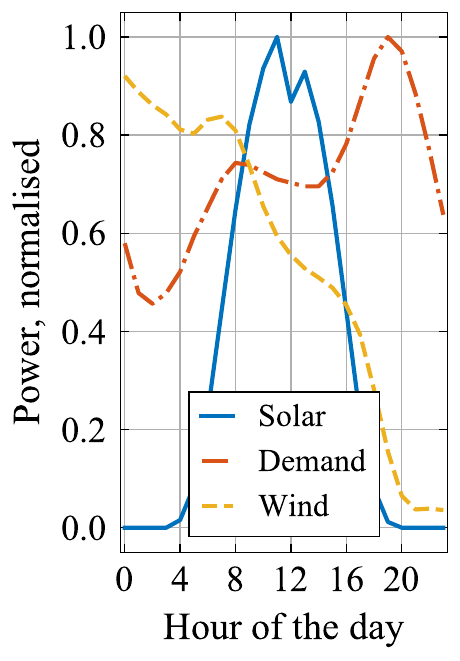}\label{f:pltDemandProfiles}}}
\caption{Two networks are used for the case studies in this work, the modified IEEE 33 bus network (a) and the UKGDS HVUG network (b), with the location of the MOP, and solar/wind generators are as indicated. Solar, wind and demand profiles used for simulations are plotted in (c).}
\label{f:network_figures}
\end{figure}

Having explored the use of the CCV to explore properties of the locus of feasible power transfers, we now consider the application of the MOP in a distribution system considering the loss reduction use-case, as described in Section~\ref{ss:formulation}. Three case studies are considered, with two of these on the IEEE 33 Bus network, and a final case study on the 75 bus UK Generic Distribution System (UKGDS) underground (UG) test system. The topology of these networks, along with the demand, wind and solar profiles, are plotted in Fig.~\ref{f:network_figures}.

\subsubsection{Case Study 1, 33 Bus Network, 750~kVA MOP}

The first case that is considered is the 33 Bus network with a 750~kVA MOP. In the first instance, three designs are considered, the conventional SOP $ \UaFx(4) $, a 3-converter Golden design $ \Ga(3) $, and the Idealised MOP $ \Oa $. These three designs are shown to clearly demonstrate how the MOP's ability to reconfigure power converters increases network performance.

\begin{figure}\centering
\subfloat[System losses (from \eqref{e:obj_val}) ]{\includegraphics[width=0.195\textwidth]{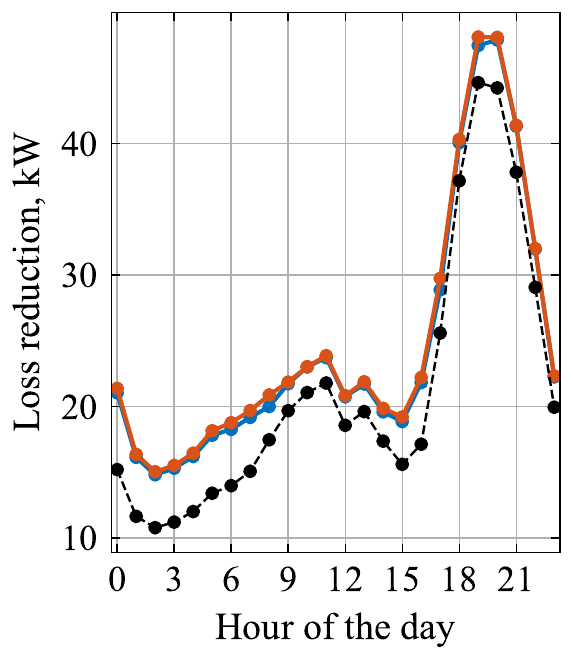}\label{f:pltCaseStudies_lossChange}}
~
\subfloat[Loss reduction ratio]{\includegraphics[width=0.26\textwidth]{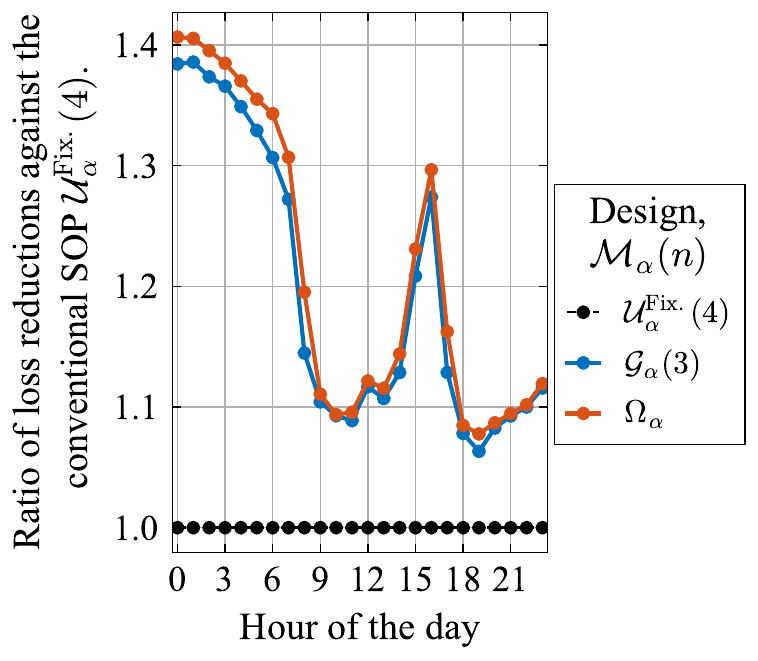}\label{f:pltCaseStudies_lossRatio}}
\caption{The reduction in total system losses (a) and ratio of this loss reduction as compared to the conventional SOP $ \UaFx $ (b) for three designs.}
\label{f:losses}
\end{figure}

Firstly, Fig.~\ref{f:losses} plots both the loss reduction for the three designs (considering both converter and network losses), and then the ratio of the hourly losses for the three designs as compared to the conventional SOP $ \UaFx $. It can be observed that all three designs show the same general pattern of loss reduction, with the highest loss reduction observed during the evening peak. The Idealised design $ \Oa $ increases the loss reduction by 40\% around 1am (Fig.~\ref{f:pltCaseStudies_lossRatio}), with similar performance from the $ \Ga(3) $ design. In fact, despite it only having three converters, the $ \Ga(3) $ design achieves 92\% of the benefits that the Idealised design $ \Oa $.

\begin{figure}\centering
\subfloat[Conventional SOP]{\includegraphics[width=0.165\textwidth]{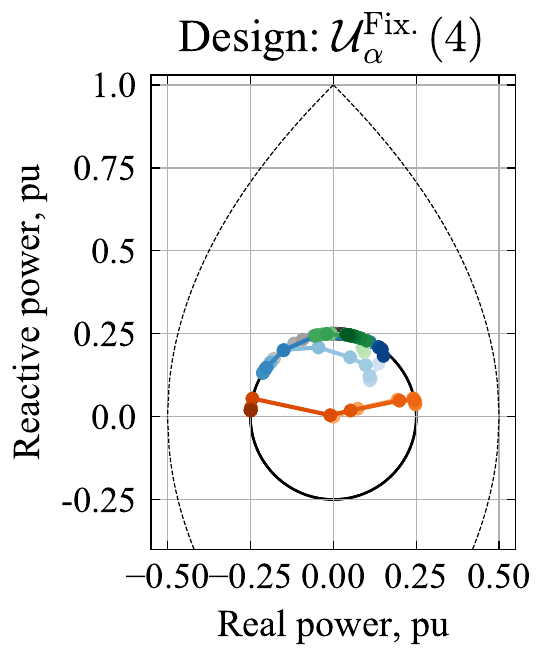}\label{f:pltComplexPowers_0}}
\subfloat[$\Ga(3)$ MOP]{\includegraphics[width=0.132\textwidth]{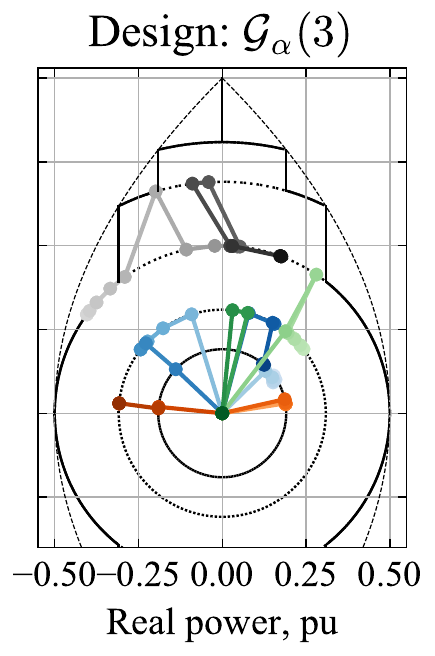}\label{f:pltComplexPowers_1}}
\subfloat[Idealised MOP]{\includegraphics[width=0.2\textwidth]{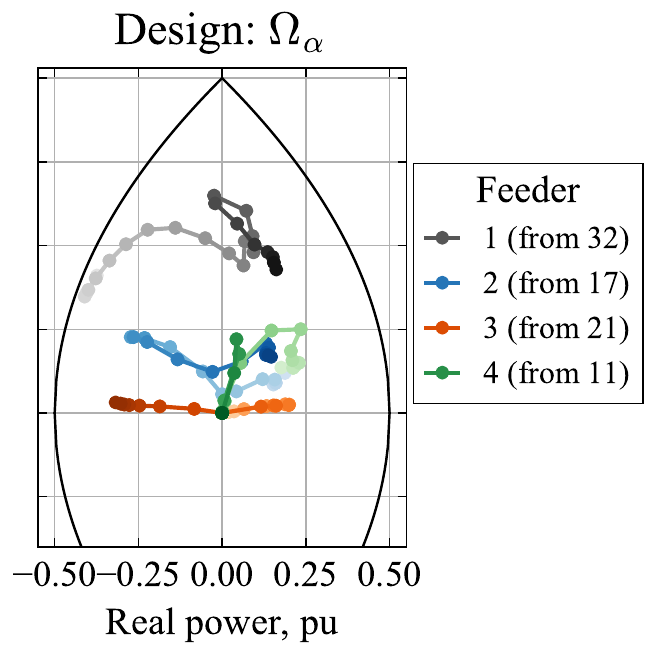}\label{f:pltComplexPowers_2}}
\caption{The optimal real and reactive power transfers for the conventional SOP $ \UaFx(4) $ (a), the 3-converter Golden MOP $ \Ga(3) $, and the Idealised MOP $ \Oa $. The four colours indicate the injections for each feeder; the brightness indicates the time of day, with earlier hours of the day plotted in a lighter shade. The solid outer lines indicate the maximum per-feeder powers that can be transferred for the three designs.}
\label{f:pltComplexPowers}
\end{figure}

The reason for the improved performance can be explained by considering the powers transferred by the MOP, as indicated in Fig.~\ref{f:pltComplexPowers}. This figure plots the real and reactive powers across the day, with each feeder represented by a colour, and the shades of each hue indicating the time of day (lightest shade is earliest in the day). Fig.~\ref{f:pltComplexPowers_2} plots the injections for the Idealised MOP $ \Oa $. This shows that Feeder 1 injects real power into the MOP from bus 32 during the morning (due to high wind, shown in Fig.~\ref{f:network_figures}); that Feeder 2 injects real power from bus 17 into the MOP around midday (due to high solar output); and then finally that Feeder 3 injects power from bus 21 in the late evening (due high demand and low solar and wind output). Additionally, reactive power is injected throughout the day to counter loads' inductive power factor.

It is immediately apparent from Fig.~\ref{f:pltComplexPowers} that the feasibility chart and realised power transfers of the Golden design $ \Ga(3) $ are much closer to operation of the Idealised design $ \Oa $ than the conventional SOP $ \UaFx(4) $, due to the greatly expanded capability chart. Qualitatively, however, it can be observed that the realisable Golden design $ \Ga(3) $ has discrete transitions between hours, as the power converters switch between feeders. In contrast, the ac/dc converters in the conventional SOP $ \UaFx $ are saturated at 25\% of the total capacity for much of the day.

\begin{figure}\centering
\subfloat[Feeder 1 (case 1)]{\includegraphics[width=0.195\textwidth]{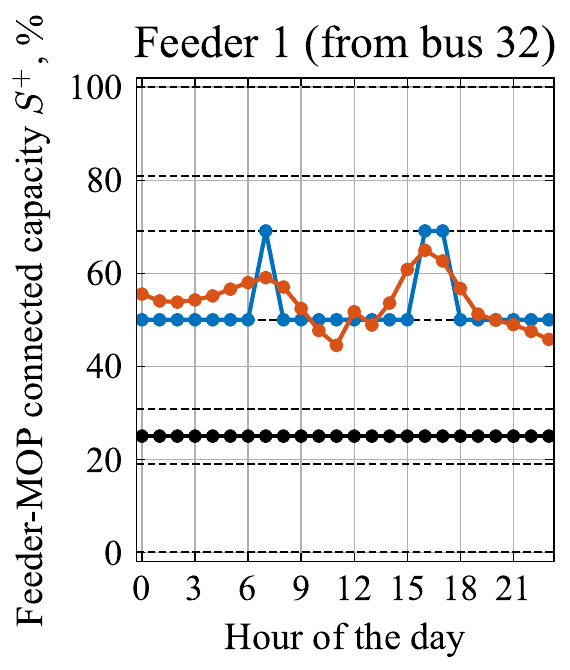}\label{f:pltCaseStudies_capconfdr_0}}
~
\subfloat[Feeder 4 (case 1)]{\includegraphics[width=0.26\textwidth]{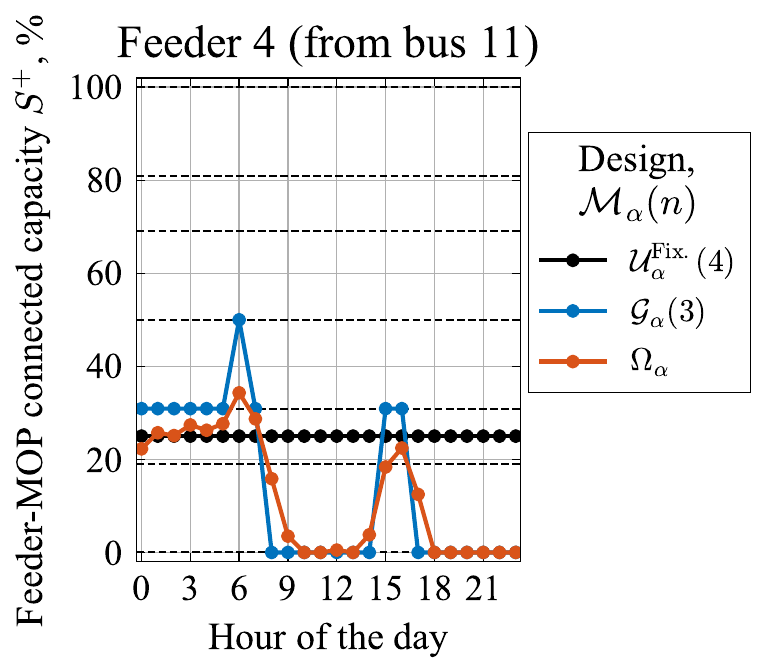}\label{f:pltCaseStudies_capconfdr_3}}
\caption{The Feeder-MOP connected capacity $ S^{+} $ (from \eqref{e:conn_capacity_feeders}) for Feeder 1 (a) and Feeder 4 (b), considering three MOP designs. Dashed horizontal lines indicating the feasible capacity combinations that are possible to connect to each feeder for the Golden MOP design $ \Ga(3) $.}
\label{f:pltCaseStudies_capconfdr}
\end{figure}

These transitions can be seen even more clearly by considering the MOP-feeder capacity $ S^{+} $ (from \eqref{e:conn_capacity_feeders}), as plotted in Fig.~\ref{f:pltCaseStudies_capconfdr}. It can be seen that 40\% or more of the ac/dc converter capacity is allocated to Feeder 1 (connected to bus 32) throughout the day for both the Golden $ \Ga(3) $ and Idealised $ \Oa $ MOP designs. In contrast, for Feeder 4 (connected to bus 11), it is optimal to assign no ac/dc converter capacity for the period around midday and in the evening, with it instead being preferable for that capacity to be allocated elsewhere. In contrast, the conventional SOP $ \UaFx $ has 25\% of the total ac/dc capacity connected to both feeders throughout the day, with that design being unable to reallocate depending on need.

As well as considering the performance of the Golden design, it is also possible to use the framework described in Section~\ref{ss:opf_mop_benefits} to evaluate the relative benefits $ \mu $ of any MOP design. Table~\ref{t:tblDesignComparison_tsg_tsg33_1_updated} summarises the results for a range of designs, comparing the performance of the three sizing strategies $ \Ua,\,\Ba,\Ga $ (as described in Section~\ref{ss:mop_sizing_strategies}) for different converter numbers $ n $. It can be seen that all of the sizing strategies show a relative benefit $ \mu $ of 10\% or more for 3 converter designs, with the relative performance $ \eta $ indicating that the six-converter Bisection design $ \Ba(6) $ achieves 99\% of the benefit provided by the Idealised MOP $ \Oa $. For smaller numbers of converters, the Golden sizing strategy performs best, whilst for larger numbers of converters, the Bisection sizing shows better performance.

\begin{table}
\input{tblDesignComparison_tsg_tsg33_1_updated}
\label{t:tblDesignComparison_tsg_tsg33_1_updated}
\end{table}

\subsubsection{Cases 2 and 3} Case 2 and 3 have 100~kVA of ac/dc converter capacity connected. These cases are considered with the Golden sizing strategy $ \Ga $ only, as it showed good performance for Case 1. Case 2 is again based on the IEEE 33 Bus network with a 4-terminal MOP, whilst Case 3 is based on the 75 bus UKGDS HVUG network with a 2-terminal MOP. Performance metrics for these two cases are reported in Table~\ref{t:tblDesignComparison_tsg_tsg33sml_1_updated} and Table~\ref{t:tblDesignComparison_tsg_tsgGds_1_updated} respectively.

For both of these cases, it can be seen that the benefit $ \mu $ of the reconfigurable MOP is substantially higher even when the number of converters $ n $ is only 2, and that the relative performance $ \eta $ is as high as 99\% even when there are just $ n=4 $ converters. This supports the observation that can be made from Table~\ref{t:tblDesignComparison_tsg_tsg33_1_updated} that excellent network performance $ \eta $ can be achieved even with a small number of converters.

\begin{table}
\input{tblDesignComparison_tsg_tsg33sml_1_updated}
\label{t:tblDesignComparison_tsg_tsg33sml_1_updated}
\end{table}

\begin{table}
\input{tblDesignComparison_tsg_tsgGds_1_updated}
\label{t:tblDesignComparison_tsg_tsgGds_1_updated}
\end{table}

\begin{figure}\centering
\subfloat[33 Bus, $ g^{*}=95.4 $ kWh/day]{\includegraphics[width=0.21\textwidth]{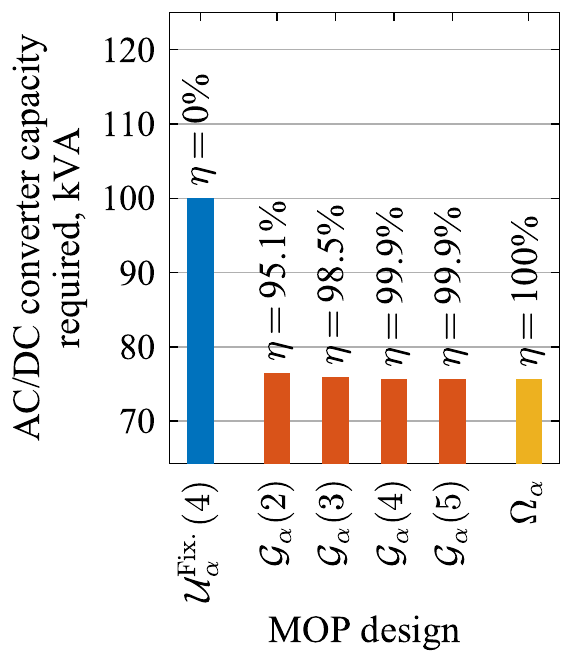}\label{f:pltMopSizes_tsg33sml}}
~
\subfloat[HVUG, $ g^{*}=5.48 $ kWh/day]{\includegraphics[width=0.21\textwidth]{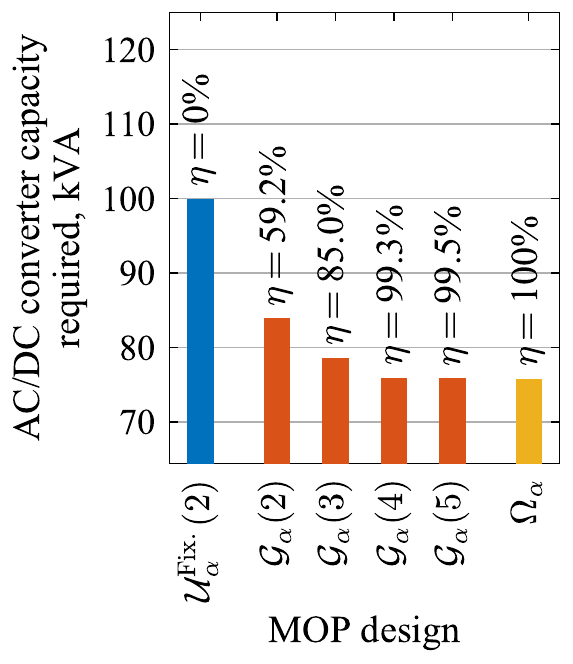}\label{f:pltMopSizes_tsgGds}}
\caption{Equivalent network benefits $ g^{*} $ \eqref{e:tot_opf_benefit} can be achieved with a reduced ac/dc converter capacity requirement for both Case 2 (33 Bus network, (a)) and Case 3 (UKGDS HVUG network, (b)).}
\label{f:pltMopSizes}
\end{figure}

Both of these cases show close to 30\% relative benefits $ \mu $. If it is assumed that the service level (i.e., loss reduction $g^{*} $) achieved by the 100~kVA conventional SOP is appropriate for the network, the reduction in ac/dc converter capacity required to achieve this service can be determined using root-finding methods. Fig.~\ref{f:pltMopSizes} plots the results of applying the Secant root-finding method to to 0.1\% relative accuracy. The $ \Ga(5) $ design requires 78.9~kVA and 75.7~kVA ac/dc converter capacity for Case 2 (33 Bus) and Case 3 (HVUG) respectively; likewise, the Idealised design $ \Oa $ requires 78.8~kVA and 75.7~kVA respectively. Assuming a price of \$100/kVA for ac/dc converter capacity \cite{huber2017applicability}, these 21\% and 24\% savings result in a cost reduction from \$10k to \$7.9k and \$7.6k, respectively.

\subsection{Discussion}\label{ss:case_discussion}

The main driver for this work has been the high unit cost of power electronics, motivating methods to reduce the capacity of ac/dc converters that are required to provide a given network service (or conversely, methods of increasing converter flexibility). However, cost is not the only factor that DSOs consider when selecting power electronics-based solutions--instead, there are a number of further important design criteria to consider if a converter topology is to be attractive for DSOs \cite{mcgibbon2020power}. It is interesting to note that the MOP philosophy is well-suited to address many of these criteria. For example, to allow maintenance on the ac/dc legs, the MOP could be designed with a degree of redundancy, so that an individual ac/dc converter could be switched out for a period for maintenance whilst the remaining ac/dc converters maintain a base level of service. Similarly, increasing the number of parallel ac/dc converters increases the reliability of the power electronics in the case of isolated failures, although it is worth noting that electromechanical switches (multiplexers) also introduce an additional failure mode. Future work could explore designs that address these supplementary design criteria directly whilst keeping the required ac/dc converter capacity low. 

\DM{Voltage source converters are highly flexible in their potential support of networks in post-fault conditions (or during maintenance) \cite{cao2016operating}. For example, it has been proposed that SOPs are operated alongside network reconfiguration technologies to support network resilience \cite{yang2022resilience} or reliability \cite{zhang2022multiple}, and the proposed MOP could be integrated into these approaches. The switches in the multiplexer, however, add operational complexity to changes in operating state of the device (e.g., if the topology of the converter needs to be changed to inject powers in a post-fault system state). These multiplexers will therefore need to be specified to meet necessary operating criteria during both normal operation and under emergency conditions. For example, if seamless converter reconfiguration is required over a few milliseconds, then it may be necessary to use solid-state static transfer switches. Alternatively, this reconfiguration is acceptable over the course of hundreds of milliseconds to seconds, then electromechanical contactors or disconnector switches may be sufficient \cite{fuad2020soft}. Future work could consider the advantages and disadvantages of each approach considering relevant grid codes.}

It is also interesting to consider unbalanced generalizations of the balanced MOP considered in the paper. It is implicit in this work that the sizing of each ac/dc converters is on a three-phase basis, as it has been assumed that power flows through the MOP are balanced. There is no reason that the approach could not be applied to unbalanced SOPs as well \cite{li2017optimal}. Asymmetric sizing of the ac/dc converter legs per-phase could add further flexibility to the `phase-changing SOP' approach explored in \cite{lou2020new}. As customers connected to LV systems often have very variable demand profiles, this flexibility could yield significant increases in performance. \DM{Future work could also explore the potential of the multiplexing approach for other converter topologies, such as shunt-series unified power flow controllers (UPFCs) \cite{maza2012voltage,bloemink2013benefits}, which have also shown potential benefits in distribution network applications.}

It is also worthwhile noting that in such an unbalanced case that two pairs of unbalanced three-phase LV distribution cables meeting at a normally open point would appear as a system with $m=6$ `feeders'. As noted in Section~\ref{ss:mop_use_cases}, such a system would have an 11-dimensional CCV. Whist Monte Carlo Integration methods can work in such cases, the time to converge can be very slow due to the `curse of dimensionality', with the number of samples required increasing very rapidly with dimension \cite{press2007numerical}[Ch. 4.8]. Future works could explore methods of evaluating CCVs using more efficient techniques (e.g., importance subsampling \cite{press2007numerical}[Ch. 7]) to unlock more sophisticated analysis of CCVs in higher dimensions.

\section{Conclusions}\label{s:6_conclusions}

This paper has proposed the general Multiplexed Soft Open Point, a device that connects each of $n$ ac/dc power converters to $m$ feeders through a bank of multiplexers. It has been shown that equivalent network performance can be achieved using a smaller per-unit capacity of power converters. Assuming that the per-unit cost of capacity of the individual ac/dc converters is much higher than electromechanical switches, this yields a reduction in device cost.

MOP performance has been evaluated via two categories of complementary metrics: through network-agnostic Capability Chart Volumes (CCVs), based on a quantification of the size of the set of feasible power transfers, and through network-aware relative benefits and performance improvement metrics, considering a loss reduction use-case. The proposed use of converter sizing strategies admits an idealised MOP design that can provide an upper bound for these metrics, with the mathematical convenience of modelling this idealised device resulting in both a useful and practical construct. 

Analytic and computational methods have been used to show the value of the CCV can be increased by more than a factor of two, with a loss reduction use-case increasing performance by 30\% (or conversely, reducing the necessary ac/dc converter capacity to achieve equivalent performance by 24\%). Three sizing strategies are explored, with realisable designs demonstrated to achieve a relative performance of 99.9\% as compared to the upper bound based on the idealised MOP design. Further challenges around the real-world application of the device, such as reliability of the electromechanical multiplexers, have been discussed.

The unit cost of power electronics capacity is high, and so it is imperative that the power converters that are installed are operated with the highest possible utilization. It is concluded that the MOP philosophy yields flexible SOP designs that consistently outperform conventional hard-wired SOP approaches, thereby showing potential to provide much-needed network services and capacity for distribution networks.

{\appendices
\section*{Appendix--Evaluating the CCV for a (2, 3) MOP}
The value of $ \ccvpq $ \eqref{e:capability_chart_volume} will be calculated for the Golden design $ \Ga(3) $, as required for the results in Section~\ref{ss:case_ccvs}. The evaluation of $ \ccvpq $ for $ \Ba(3),\,\Ua(3) $ is analogous, and so for brevity the method is not repeated. 

First, denote the integral $W(\theta,\gamma)$ \cite[pp. 282]{gradshteyn2007table}
\begin{align}\label{e:wacky_integral}
W(\theta ,\,\gamma) &= \int_{0}^{\gamma} \sqrt{(\theta ^{2} - x^{2})(\gamma^{2} - x^{2})}\, dx\\
		&= \dfrac{\theta  \left ( \theta E(\tau) - (\theta ^{2}-\gamma^{2})K(\tau)\right )}{3} \,, \; \tau=\left (\dfrac{\gamma}{\theta }\right )^{2},
\end{align}
where $\theta \geq \gamma \geq 0$, and $ E,\,K $ are elliptic integrals of the first and second kind. 

The approach used to determine \eqref{e:capability_chart_volume} is to determine the area in the $ Q[1],\,Q[2] $ plane as a function of $ P $, reducing the three-dimensional volume integral to an integration of a single variable. Fig.~\ref{f:appendix_proof} plots the area $ A $ in the lower half of the first quadrant, showing three regions $ r_{1},\,r_{2},\,r_{3} $ that are used to calculate the total area depending on $ |P| $. When real powers transfers $|P|$ are large, there is only one feasible configuration of the three ac/dc converters; real power transfers smaller than the smallest converter $ \alpha[3] $ admits three configurations.

\begin{figure}
\centering
\includegraphics[width=0.48\textwidth]{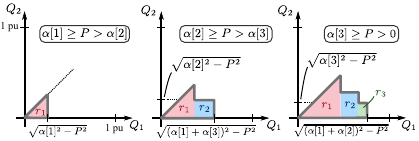}
\caption{Feasible combinations of reactive power in the lower half of the first quadrant for large real power transfers (greater than $ \alpha[2] $, left), intermediate power transfers (between $ \alpha[2] $ and $ \alpha[3] $, centre) and small transfers (less than $\alpha[3]$), assuming elements of $ \alpha $ are ordered in decreasing size.}
\label{f:appendix_proof}
\end{figure}

The integral of the area $r_{1}$ is
\begin{equation}\label{e:r1}
R_{1} = \dfrac{1}{2}\int_{0}^{\alpha[1]=1/2} \left( \dfrac{1}{2}\right)^{2} - P^{2} \,dP
\end{equation}
which has value 1/24. The integral of areas $ r_{2},\,r_{3} $ are

\begin{align}\label{e:r2}
&\begin{aligned}
&R_{2} = \int_{0}^{\alpha[2]} \sqrt{\left ( (\alpha[1]+\alpha[3])^{2} - P^{2}\right ) \left (\alpha[2]^{2} - P^{2} \right ) }\\
& \qquad \qquad - \sqrt{\left ( \alpha[1]^{2} - P^{2}\right ) \left (\alpha[2]^{2} - P^{2} \right ) }\; dP
\end{aligned}\\
&\hspace{0.7em} = W\left ( (\alpha[1]+\alpha[3]),\alpha[2]\right ) - W(\alpha[1],\alpha[2])\,,\\
&\hspace{-0.5em}R_{3} = W ( (\alpha[1]+\alpha[2]),\alpha[3]) - W ((\alpha[1]+\alpha[3]),\alpha[3] )\,.\label{e:r_3}
\end{align}
Considering the whole $ Q[1],\,Q[2] $ plane, and both positive and negative real power transfers, yields
\begin{equation}\label{e:result}
\mathrm{CCV} = 16\sqrt{2}(R_{1}+R_{2}+R_{3})\,.
\end{equation}

}

\input{tsg2022.bbl}

\bibliographystyle{IEEEtran}

\vfill

\end{document}

%% file: tables/analytic2feeder.tex
\centering
\caption{CCVs for $m=2$ feeders (as defined in Section~\ref{ss:mop_use_cases}), varying the number of ac/dc converters $n$ and sizing strategy $ \Ma $.}

\begin{tabular}{llllll}
\toprule

\multirow{2}*{\vspace{-1.5em} \begin{tabular}[x]{@{}l@{}}Sizing\\strategy,\\$ \mathcal{M}_{\alpha} $ \end{tabular}} & $ \multirow{2}*{\vspace{-1.5em} $ n $} $ & \multicolumn{4}{c}{Capability Chart Volume (CCV) mode ($m=2$ feeders)}\\
  \cmidrule(l{0.6em}r{0.9em}){3-6}
&  & MPT, kW & UPF, kW & \tcell{STATCOM,\\kVAr$^{2}$} & \tcell{PQ,\\kWkVAr$ ^{2} $} \\

\midrule

$\UaFx$ & 2 & $ \frac{1}{2} $  	& $ \sqrt{2} $ 	& 1 	& 0.943\vspace{1.0em} \\

$ \Ua $ & 3 & $\frac{1}{3}$ & $ \frac{2\sqrt{2}}{3} $ 	& $ \frac{4}{3} $ 	& 0.995\vspace{0.3em} \\

$ \Ba $ & 3 & $\frac{1}{2}$ & $ \sqrt{2} $ 	& $\frac{3}{2}$ 	& 1.227\vspace{0.3em} \\

$ \Ga $ & 3 & $\frac{1}{2}$ & $ \sqrt{2} $ 	& 1.652 	& 1.357 \vspace{1.0em} \\

$ \Oa $ & - & $\frac{1}{2}$ & $ \sqrt{2} $ & 2 & 1.744 \\

\bottomrule
\end{tabular}

%% file: tables/tblDesignComparison_tsg_tsg33_1_updated.tex
\centering
\caption{Effect of sizing strategies $ \Ma $ and number of AC/DC converters $ n $ for the 33 Bus case study with a 750 kVA MOP, considering relative benefit $ \mu$ \eqref{e:rel_ben_defn} and performance $ \eta $ \eqref{e:rel_perf_defn}.}\label{t:tblDesignComparison_tsg_tsg33_1}
\begin{tabular}{lllll}
\toprule
\begin{tabular}[x]{@{}l@{}}Design,\\$\Ma$\end{tabular} & \begin{tabular}[x]{@{}l@{}}Number of\\converters, $n$\end{tabular} & \begin{tabular}[x]{@{}l@{}}Loss reduction,\\$g^{*}$, kWh/day\end{tabular} & \begin{tabular}[x]{@{}l@{}}Relative\\Benefit,\\$\mu$, \%\end{tabular} & \begin{tabular}[x]{@{}l@{}}Relative\\Performance,\\$\eta$, \%\end{tabular} \\
\midrule
$\mathcal{U}_{\alpha}^{\mathrm{Fix.}}$ & 4 & 509.9 & 0\% & 0\% \vspace{0.5em} \\
$\mathcal{U}_{\alpha}$ & 2 & 520.7 & 2.1\% & 12.5\% \\
$\mathcal{U}_{\alpha}$ & 3 & 567.8 & 11.4\% & 66.7\% \\
$\mathcal{U}_{\alpha}$ & 4 & 584.9 & 14.7\% & 86.3\% \\
$\mathcal{U}_{\alpha}$ & 5 & 586.9 & 15.1\% & 88.6\% \\
$\mathcal{U}_{\alpha}$ & 6 & 590.0 & 15.7\% & 92.2\% \vspace{0.5em} \\
$\mathcal{B}_{\alpha}$ & 3 & 584.9 & 14.7\% & 86.3\% \\
$\mathcal{B}_{\alpha}$ & 4 & 590.8 & 15.9\% & 93.1\% \\
$\mathcal{B}_{\alpha}$ & 5 & 595.2 & 16.7\% & 98.1\% \\
$\mathcal{B}_{\alpha}$ & 6 & 595.9 & 16.9\% & 99.0\% \vspace{0.5em} \\
$\mathcal{G}_{\alpha}$ & 3 & 589.5 & 15.6\% & 91.7\% \\
$\mathcal{G}_{\alpha}$ & 4 & 592.5 & 16.2\% & 95.1\% \\
$\mathcal{G}_{\alpha}$ & 5 & 593.6 & 16.4\% & 96.4\% \\
$\mathcal{G}_{\alpha}$ & 6 & 593.7 & 16.4\% & 96.4\% \vspace{0.5em} \\
$\Omega_{\alpha}$ & - & 596.8 & 17.0\% & 100\% \\
\bottomrule
\end{tabular}

%% file: tables/tblDesignComparison_tsg_tsg33sml_1_updated.tex
\centering
\caption{Comparing the relative benefit $ \mu $ \eqref{e:rel_ben_defn} and relative performance $\eta$ \eqref{e:rel_perf_defn} for the 33 Bus network with a 100 kVA MOP.}\label{t:tblDesignComparison_tsg_tsg33sml_1}
\begin{tabular}{lllll}
\toprule
\begin{tabular}[x]{@{}l@{}}Design,\\$\mathcal{M}_{\alpha}$\end{tabular} & \begin{tabular}[x]{@{}l@{}}Number of\\converters, $n$\end{tabular} & \begin{tabular}[x]{@{}l@{}}Loss reduction,\\$g^{*}$, kWh/day\end{tabular} & \begin{tabular}[x]{@{}l@{}}Relative\\Benefit,\\$\mu$, \%\end{tabular} & \begin{tabular}[x]{@{}l@{}}Relative\\Performance,\\$\eta$, \%\end{tabular} \\
\midrule
$\mathcal{U}_{\alpha}^{\mathrm{Fix.}}$ & 4 & 95.4 & 0\% & 0\% \vspace{0.5em}\\
$\mathcal{G}_{\alpha}$ & 2 & 122.5 & 28.4\% & 95.1\% \\
$\mathcal{G}_{\alpha}$ & 3 & 123.4 & 29.4\% & 98.5\% \\
$\mathcal{G}_{\alpha}$ & 4 & 123.8 & 29.8\% & 99.9\% \\
$\mathcal{G}_{\alpha}$ & 5 & 123.8 & 29.8\% & 99.9\%  \vspace{0.5em}\\
$\Omega_{\alpha}$ & - & 123.9 & 29.9\% & 100\% \\
\bottomrule
\end{tabular}

%% file: tables/tblDesignComparison_tsg_tsgGds_1_updated.tex
\centering
\caption{Comparing the relative benefit $ \mu $ \eqref{e:rel_ben_defn} and relative performance $\eta$ \eqref{e:rel_perf_defn} for the UKGDS HV UG network with a 100 kVA MOP.}\label{t:tblDesignComparison_tsg_tsgGds_1}
\begin{tabular}{lllll}
\toprule
\begin{tabular}[x]{@{}l@{}}Design,\\$\mathcal{M}_{\alpha}$\end{tabular} & \begin{tabular}[x]{@{}l@{}}Number of\\converters, $n$\end{tabular} & \begin{tabular}[x]{@{}l@{}}Loss reduction,\\$g^{*}$, kWh/day\end{tabular} & \begin{tabular}[x]{@{}l@{}}Relative\\Benefit,\\$\mu$, \%\end{tabular} & \begin{tabular}[x]{@{}l@{}}Relative\\Performance,\\$\eta$, \%\end{tabular} \\
\midrule
$\mathcal{U}_{\alpha}^{\mathrm{Fix.}}$ & 2 & 5.48 & 0\% & 0\% \vspace{0.5em}\\
$\mathcal{G}_{\alpha}$ & 2 & 6.41 & 16.9\% & 59.2\% \\
$\mathcal{G}_{\alpha}$ & 3 & 6.82 & 24.3\% & 85.0\% \\
$\mathcal{G}_{\alpha}$ & 4 & 7.04 & 28.4\% & 99.3\% \\
$\mathcal{G}_{\alpha}$ & 5 & 7.05 & 28.5\% & 99.5\% \vspace{0.5em}\\
$\Omega_{\alpha}$ & - & 7.05 & 28.6\% & 100\% \\
\bottomrule
\end{tabular}